\documentclass[final,3p]{elsarticle}
\usepackage{amsmath}
\usepackage{graphicx}
\usepackage{amssymb}
\usepackage{amsthm}
\usepackage{tabularx}
\usepackage{graphicx}
\usepackage{subfigure}
\usepackage{booktabs}
\usepackage[hang, small,labelfont=bf,up,textfont=it,up]{caption}
\usepackage{rotating}
\usepackage{relsize}
\usepackage{setspace}

\journal{Materials and Design}

\begin{document}

\begin{frontmatter}
\title{Thermomechanical properties of amorphous metallic tungsten-oxygen and tungsten-oxide coatings}
\author{E. Besozzi$^a$, D. Dellasega$^{a,b}$, V. Russo$^{a}$, C. Conti$^{c}$,  M. Passoni$^{a,b}$, M.G. Beghi$^a$}
\address{$^a$ Dipartimento di Energia, Politecnico di Milano, via Ponzio 34/3, I-20133, Milano,Italy}
\address{$^b$ Associazione EURATOM-ENEA, IFP-CNR, Via R. Cozzi 53, 20125 Milano, Italy}
\address{$^c$ ICVBC-CNR, Via Cozzi 54, 20125, Milano, Italy}

\begin{abstract}
Metallic amorphous tungsten-oxygen and amorphous tungsten-oxide films, deposited by Pulsed Laser Deposition, are characterized. The correlation is investigated between morphology, composition, and structure, measured by various techniques, and the mechanical properties, characterized by Brillouin Spectroscopy and the substrate curvature method. 
The stiffness of the films is correlated to the oxygen content and the mass density. The elastic moduli decrease  as the mass density decreases and the oxygen-tungsten ratio increases. A plateau region is observed around the transition between the metal-like (conductive and opaque) films and the oxide ones (non conductive and transparent). 
The compressive residual stresses, moderate stiffness and high local ductility of compact amorphous tungsten-oxide films are interesting for applications involving thermal or mechanical loads. The coefficient of thermal expansion is quite high (8.9 $\cdot$ 10$^{-6}$ K$^{-1}$), being strictly correlated to the amorphous structure and stoichiometry of the films. 
Upon thermal treatments the coatings show a quite low relaxation temperature of 450 K. Starting from 670 K, they crystallize into the $\gamma$ monoclinic phase of WO$_3$, the stiffness increasing by about 70\%. 
The measured thermomechanical properties provide a  guidance for the design of devices which include a tungsten based layer, in order to assure their mechanical integrity.
\end{abstract}
\end{frontmatter}


\section{Introduction}

\label{Introduction} Tungsten oxide (WO$_{x}$), of stoichiometric or nearly
stoichiometric composition (2.8$<$O/W$<$3.1), and amorphous
metallic tungsten-oxygen (W-O), with 2$<$O/W$<$2.8, are currently the object
of several investigations, due to their interesting functional properties.
In most cases these materials are in the form of supported films. Compact
amorphous WO$_{3}$\ films with tuned electrical and optical properties are
exploited for electrochromic devices \cite{Zheng_2011, Wang_2018,
Granqvist_1993}, as contact electrodes in advanced solar cells \cite%
{Zheng_2011, Nunez_2015, Samad_2017} and for smart windows \cite%
{Karuppasamy_2007}. Porous WO$_{3}$\ films, due to their high active
specific area, are adopted for photoelectrochemical water splitting \cite%
{Li_2014, Shin2015} and photocatalysis \cite{Deb_2008, Kikuchi}. W-O films,
with 2$<$O/W$<$2.8, have properties intermediate between those of oxides and
of amorphous metals. Their color is not silvery, but dark blue; they also
show an electrical resistivity in the range 1-100 $\Omega $\ cm \cite{yamamoto}.
Finally, metallic amorphous-like tungsten films \cite{dellasegaJAP},
 with 0.3$<$O/W$<$0.6, can be annealed in a reducing atmosphere
producing, by an alternative route, tungsten-oxide nanowires. Such nanowires
show unique electric and electrochromic properties, useful for gas sensors
and catalysis \cite{DellasegaNW, Sel}. The functional properties of these
films strictly depend on the specific morphology, structure and
stoichiometry, and have triggered a strong characterization effort. \newline
These materials are typically exploited in the form of coatings, whose
mechanical integrity is often crucial for the functional performance. In
turn, the mechanical integrity basically depends on the thermomechanical
properties. For example, compressive or tensile residual stresses affect the
behavior of the films, by mitigating or favoring crack formation. Similarly,
in high temperature applications, a significant mismatch between the
coefficients of thermal expansion of the coating and of the substrate can
induce high interface stresses, with possible coating delamination and
device failure. More specifically, in an electrochromic system the W oxide
film is part of a complex multilayer system: it is deposited on a
transparent conductor, like ITO, and faces the electrolyte, solid or liquid,
containing the ions responsible of the electrochromic effect, and can be
subject to various and very different stress states \cite{gunnar}. Moreover,
in some applications (e.g. solar-cells, thermophotovoltaic) tungsten oxide
coatings operate at temperatures above room temperature \cite{Nandi_2014};
this could induce phase transition or recrystallization, with a consequent
variation of the as-deposited properties. Although the thermomechanical
properties of tungsten based coatings can be crucial for the design of
devices which exploit them, relatively \ fewer studies have investigated the
relationship between their nanostructure, composition and mechanical
properties \cite{Carrejo_2016, Parreira_2006, Hasan_2013}. The goal of this
work is to achieve a more comprehensive understanding of the effects of
structure, morphology and chemical composition on the thermomechanical
properties of different systems of amorphous W-O and WO$_{x}$\ coatings,
providing useful results for the design of devices. We investigate amorphous
films characterized by different oxygen/tungsten ratios and morphologies. To
produce them we selected the Pulsed Laser Deposition (PLD) technique, which
allows a significant versatility in tailoring the structure, the morphology
and the O$_{2}$\ enrichment of the deposited samples \cite{Pezzoli_2015,
Baserga_2007, Bailini_2007}. 
Due to the wide variety of devices, and to the different types of substrates 
on which tungsten based coatings can be deposited, a single optimal set of 
properties cannot be identified. The compatibility of thermomechanical properties 
between the coating and the substrate is often more important than the absolute 
value for the coating alone: each application can have its own most suitable coating. 
Our investigation explores an interval of properties, with the aim of providing 
information useful to determine the most appropriate coating for each single application.

The morphology, structure and stoichiometry of
the coatings are monitored by Scanning Electron Microscopy (SEM), X-Ray
Diffraction (XRD), Raman spectroscopy and Energy Dispersive X-Ray
Spectroscopy (EDXS). The thermomechanical characterization is performed exploiting Brillouin
spectroscopy (BS) and the substrate curvature method (SC). The coupling of
these two techniques has been shown to be a powerful tool for the
characterization of nanostructured films, providing a broad, non destructive
characterization of the samples \cite{Besozzi_2016, Dellasega_2017,
Besozzi_2017}. In particular, when transparent oxide coatings are
investigated, BS can be able to derive, through the detection of surface and
bulk acoustic waves, all the elastic moduli of the films (i.e. Young Modulus
($E$), shear modulus ($G$), bulk modulus ($K$) and Poisson's ratio ($\nu $))
at the same time \cite{Ferre_2013, Kundu_Ch10}. SC, instead, can be
exploited to measure the total stresses within the films ($\sigma _{f}$),
the residual stresses ($\sigma _{res}$) and the coefficient of thermal
expansion ($CTE$, or $\alpha $) \cite{Besozzi_2017}. The thermal stability
of the coatings under high temperatures is assessed by thermal annealing
treatments at various temperatures up to 870 K. By performing SC measurements
during the annealing treatments, it is possible to derive the evolution of $%
\sigma _{f}$\ in the film. The thermally induced modifications of the
morphology, structure and properties are finally measured. 


\section{Experimental}

\label{Experimental techniques}

\subsection{Deposition, and characterization of morphology, composition 
and structure}

The coatings analyzed in this work are deposited by the PLD apparatus
described in detail in \cite{Pezzoli_2015, Baserga_2007}. A Nd:YAG laser
(pulse duration $\tau _{p}$ = 7 ns (FWHM)), operates at its 2$^{nd}$
harmonic ($\lambda $ = 532 m), focused on a W target (purity 99.9\%); the
repetition rate is 10 Hz, the laser energy $\approx $ 800 mJ and the laser
spot about 9.2 mm$^{2}$. The fluence on target is thus $\approx $ 15 J
cm$^{-2}$. W ablated species expand into a vacuum chamber (base pressure $%
\approx $ 10$^{-3}$ Pa) in presence of O$_{2}$ as background gas, with
pressure varied between 5 and 60 Pa. The films are deposited onto 300 $\mu $%
m thick Si(100) substrates, which are double side polished for SC
measurements. \newline
Morphological properties are assessed by a Zeiss Supra 40 field emission
Scanning Electron Microscope (SEM), operating at an accelerating voltage of
5 kV. The composition of the samples is determined by Energy Dispersive
X-ray Spectroscopy (EDXS) in the same SEM system, working with an
accelerating voltage of 15 kV in order to promote the excitation of K$%
_{\alpha }$ and M$_{\alpha }$ electronic shells of respectively O and W.
Each measurement is repeated three times at different points of the samples.
XRD analysis is performed by\ a Panalytical X'Pert PRO X-ray diffractometer
in $\theta /2\theta $ configuration, and by micro-Raman measurements, with a
Renishaw InVia spectrometer equipped with an Ar$^{+}$ laser ($\lambda $ =
514.5 nm), a 1800 g/mm grating and an edge filter with cut at 100 cm$^{-1}$.
The laser operates at 1 mW continuum power through a 50X objective to avoid
any local material modification. Finally, the mass density $\rho $ of the
deposited films is evaluated by combining weight measurements before and
after the deposition using a precision balance (i.e. 10$^{-4}$ g) and SEM
cross-section for thickness determination.

\subsection{Elastic moduli characterization}

\label{experimental}

The elastic moduli of the coatings are evaluated by the BS spectroscopy
setup described in \cite{Beghi_RevSci}, with a Nd:YAG
laser (continuum operation at $\approx $ 200 mW, $\lambda $ = 532 nm)
focused on the coating surface. The scattered light is collected in the
backscattering geometry without polarization analysis by a Fabry-Perot
multi-pass interferometer, operating in the tandem mode, of the Sandercock
type. \newline
In the case of sufficiently transparent materials light can be inelastically
scattered by bulk ultrasonic waves, by the \textit{elasto-optic mechanism}
(the modulation of the refractive index by a mechanical strain). The
properties of bulk waves are thus accessible \cite{Ferre_2013}. At the free
surface of solids Surface Acoustic Waves (SAWs) also exist, whose
displacement field is confined in the vicinity of the surface, and declines
with depth, the decay length being close to the wavelength. In the case of
metallic samples light cannot penetrate, and interacts only with the SAWs:
the process is mediated by the \textit{surface ripple mechanism}, i.e. the
dynamic corrugation of the surface due to the wave displacement. The
properties of SAWs are thus accessible \cite{Besozzi_2016}. Since the
properties of both bulk waves and SAWs depend on the mass density and the
elastic properties, in both cases the elastic properties can be derived as
follows. \newline
Under the assumption of a homogeneous isotropic linear elastic medium, the
elastic stiffness tensor is defined by only two independent constants, which
can be taken as $C_{11}$ and $C_{44}$. The other elastic moduli can be
expressed in terms of these two elastic constants. Such a medium supports
both longitudinal and transversal bulk waves, whose velocities $v_{L}$\ and $%
v_{T}$\ are respectively\ \cite{Kundu_Ch10}: 
\begin{equation}
v_{L}=\sqrt{\frac{C_{11}}{\rho }}  \label{vL}
\end{equation}%
\begin{equation}
v_{T}=\sqrt{\frac{C_{44}}{\rho }}  \label{vT}
\end{equation}

When bulk modes can be detected, inelastic scattering of light occurs within
the medium, where the optical wavevector is affected by the the refractive
index $n$\ of the material. The velocities $v_{L}$\ and $v_{T}$\
can be directly obtained from the frequency shifts $\Delta \omega $\ of the
bulk peaks in the Brillouin spectra as 
\begin{equation}
v_{L,T}=\frac{\Delta \omega _{L,T}\lambda _{0}}{4\pi n},  \label{bulkwaves}
\end{equation}%
where $\lambda _{0}$\ is the laser wavelength \cite{Kundu_Ch10}. In this
case $C_{11}$\ and $C_{44}$\ are obtained in a staightforward way from eqs. %
\ref{vL} and \ref{vT}, if both the refractive index and the mass density of
the medium are known. As it will be discussed in section \ref{result2}, $n$
will be estimated directly from Brillouin spectra.

Surface waves jointly depend on $C_{11}$ and $C_{44}$. If only SAWs are
detected, a suitable procedure must be adopted in order to derive them,
which is described in detail in \cite{Besozzi_2016}. The spectra are
recorded at different incidence angles $\theta $, obtaining the experimental
dispersion relation of the modes as function of $\theta $. Only the
component of wave vector parallel to the surface is relevant. This
component depends on the incidence angle $\theta $, but not on the
refractive index. The velocities \ of SAWs are obtained from the frequency
shifts in the spectrum, without needing the value of $n$, as \cite%
{Kundu_Ch10}: 
\begin{equation}
v_{SAW}=\frac{\Delta \omega _{R}\lambda _{0}}{4\pi \sin \theta }
\end{equation}

Theoretical dispersion relations can be computed by solving the
Christoffel's secular equation for an equivalent homogeneous system under
the isotropic assumption. A least squares minimization can therefore be performed,
between the computed dispersion relations and the measured ones. The
minimization is performed with $C_{11}$ and $C_{44}$ as the only free
parameters, for a fixed value of $\rho $, obtaining the most probable
estimates for $C_{11}$ and $C_{44}$. The number of the experimentally observed
modes determines, in turn, the accuracy in the determination of the elastic
properties of the films \cite{Besozzi_2016}. \newline

\subsection{Residual stresses, CTE and annealing treatments}

\label{experimental2} Residual stresses are measured by an optical
implementation of the SC method.\ An ad-hoc developed experimental setup,
fully described in \cite{Besozzi_2017}, exploits a set of parallel laser
beams to probe the curvature radius of the coating-substrate system. The
scanned area is $\approx $ 1 cm$^{2}$ and the laser beams strike on the
uncoated substrate surface before being collected by a high frame rate
camera. Residual stresses are derived by measuring the variation of the
substrate curvature before and after film deposition. According to Stoney's
approximation, for a thin supported film the residual stress $\sigma _{res}$
can be computed for the wafer curvature as \cite{Stoney}: 
\begin{equation}
\sigma _{res}=\frac{E_{s}}{1-\nu _{s}}\frac{t_{s}^{2}}{t_{f}}\frac{1}{6R_{c}}
\label{Stoney}
\end{equation}%
where $E_{s}$ and $\nu _{s}$ are the Young modulus and Poisson's ration of
the substrate, $t_{f}$ and $t_{s}$\ are the thicknesses of the film and the
substrate and $R_{c}$ is the curvature radius of the system. In this case, $R_{c}
$ is measured multiple times on the same sample varying the probed position.
\newline
Tests are performed in measurement chamber, equipped with a resistive heater stage
for measurements at high temperatures, 
temperature being measured by a thermocouple placed beneath the sample. 
High temperature tests are performed in vacuum (i.e. base pressure of 
$\approx $ 10$^{-4}$ Pa), in order to in principle avoid any modification of
the composition. \newline 
The thermal stress evolution, and the CTE, are obtained by monitoring the
substrate curvature change of the film-substrate system during fast heating
ramps ($\approx $ 50 K min$^{-1}$), for a fixed position of the laser
beams. The stress is again computed by equation \ref{Stoney}; under the
assumption of uniform material temperature the thermal stress can be
computed as: 
\begin{equation}
\sigma _{f}=\frac{E_{f}}{1-\nu _{f}}(\alpha _{f}-\alpha _{s})\Delta T
\label{StoneyTh}
\end{equation}%
From equations \ref{Stoney} and \ref{StoneyTh} the CTE of the film, $\alpha
_{f}$, can be derived, if the CTE of the substrate, $\alpha _{s}$, and the elastic moduli of the film are known.
The performances of the setup and more details on the measurements procedure
can be found in \cite{Besozzi_2017}.

Standard thermal annealing treatments are performed in the same apparatus at
temperatures between room temperature (RT) and 870 K. All treatments are
performed with a dwell annealing time of 1 h, and heating ramps set to $%
\approx $ 3 - 5 K min$^{-1}$. The substrate curvature, measured by $R_{c}$,
is monitored during heating and cooling.


\section{Results and discussion}

\label{Results and discussion}

\subsection{Morphology, composition and structure of amorphous W-O and WO$_{x}$ coatings}

\begin{figure*}[h!]
\centering
\includegraphics[width = \columnwidth]{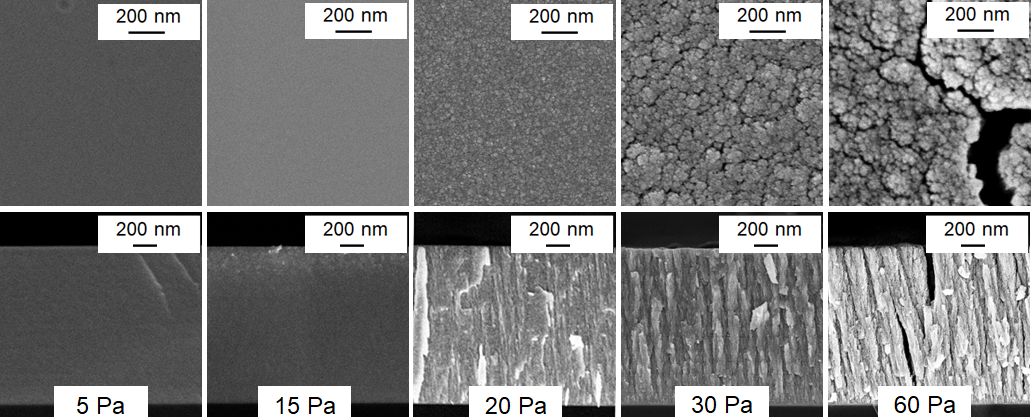}
\caption{SEM top view and cross-section images of tungsten oxide coatings
deposited by PLD at different O$_{2}$ background pressures.}
\label{SEM_asdep}
\end{figure*}

A detailed description of the growth process of these coatings by PLD can be
found elsewhere \cite{Pezzoli_2015, Baserga_2007}. 
Plane views and cross-section SEM images of the analyzed samples are shown
in figure \ref{SEM_asdep}. Samples produced at O$_{2}$ pressure below 20 Pa
are characterized by a compact and homogeneous structure. At 20 Pa a compact
nanostructured morphology appears. At higher pressures, instead, an open
porous morphology prevails. As it can be seen, at 30 Pa and 60 Pa
the pressure is sufficiently high to start promoting \textit{cauliflower}
growth. These different morphologies are related to distinct growth
mechanisms \cite{Pezzoli_2015, Baserga_2007}, strictly correlated to the
expansion dynamics of the plasma plume during deposition. Low O$_{2}$
pressures promote atom-by-atom deposition, that results in the growth of
compact films, while high O$_{2}$ pressures favors clusters formation inside
the plume, so a porous morphology. \newline
The O/W ratio, assessed by EDXS analysis, is in turn affected by the
deposition pressure. The ratios are summarized in table \ref{tab_n}. As it
can be seen, we have O/W $\approx $ 2.1 at 5 Pa, increasing to $\approx $
2.6 at 15 Pa. Above 20 Pa, instead, the films are almost stoichiometric WO$%
_{3}$. This can be associated again to an increase of interaction
probability between W and O inside the plasma plume at sufficiently high O$%
_{2}$ pressures. \newline

\begin{figure}[h!]
\centering
\includegraphics[width = 0.5\columnwidth]{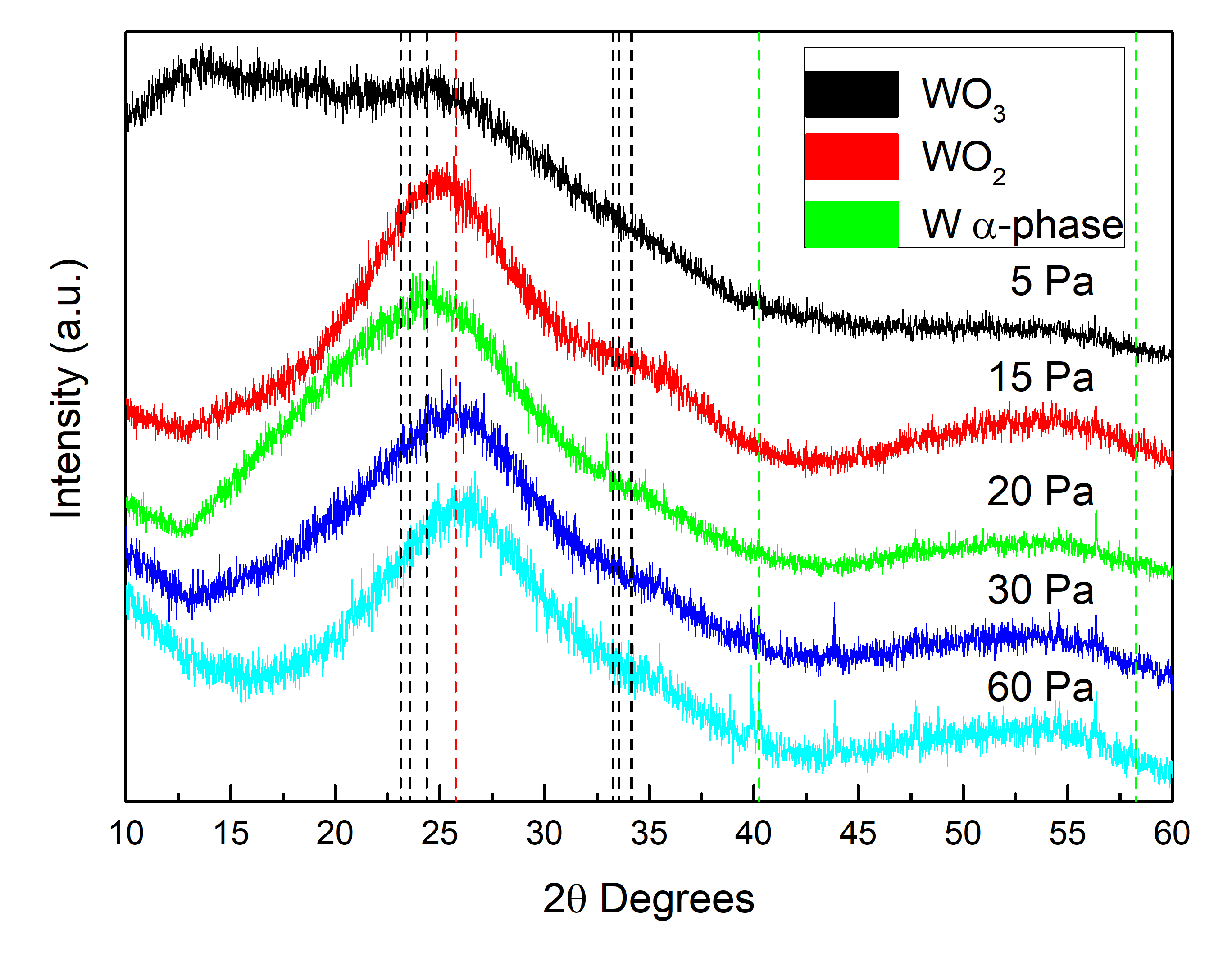}
\caption{XRD analysis of tungsten oxide coatings deposited at different O$_2$
background pressures. The shoulder at low angles for the 5 Pa sample is most
 probably an artefact due to a different sample holder.}
\label{XRD}
\end{figure}

In figure \ref{XRD} the XRD analysis of the W-O films deposited at different
O$_{2}$ pressures is shown. For all the W-O films deposited in this oxygen
pressure range the spectra exhibit a broad band around 26$%
{{}^\circ}%
$ with a small shoulder at about 35$%
{{}^\circ}%
$. The position of the main band is close to that of WO$_{3}$ and WO$_{2}$
crystalline peaks. None of the spectra contains peaks or bands due to $%
\alpha $-W or $\beta $-W metallic crystals. It is worth noting that the
present amorphous system is different from the amorphous-like W obtained by
PLD using a He atmosphere. In that case amorphous-like W exhibits instead a
broad band centered around 40$%
{{}^\circ}%
$, which is the (110) peak of $\alpha $-W \cite{dellasegaJAP}. The formation of
these two different amorphous structures related to the W-O systems had been
already found for magnetron sputtered deposits \cite{Parreira_2006}. In
particular, the films obtained at all the pressures investigated in this
work have the structure which has been called 'quasi amorphous' \cite{Parreira_2006}.
\newline
However, a semi quantitative resistance test discriminates the films
deposited at 5 Pa and 15 Pa, which are conductors, from the films deposited
at 20 Pa and above, which are insulators (resistance $>100M\Omega $).

\begin{figure}[h!]
\centering
\includegraphics[width = 0.5\columnwidth]{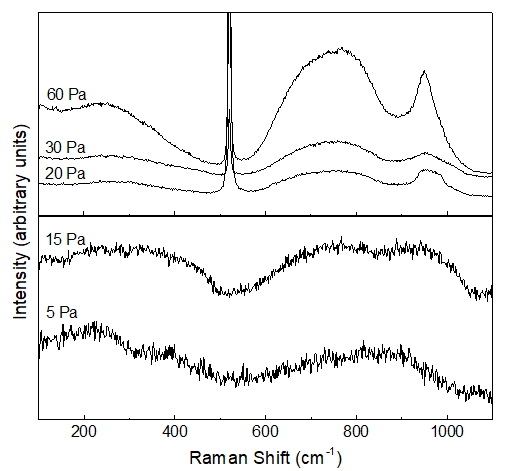}
\caption{Raman spectra of tungsten oxide coatings deposited at different O$%
_2 $ background pressures. The Si substrate peaks are at 521 cm$^{-1}$ and
at 960 cm$^{-1}$.}
\label{Raman_asdep}
\end{figure}

Thesefindings are consistent with the outcome of Raman spectroscopy, reported in
figure \ref{Raman_asdep}. All the spectra show two broad bands, a low
frequency band in the range of 100 - 500 cm$^{-1}$, associated to the O-W-O
bending modes, and a high frequency band in the range of 600 - 900 cm$^{-1}$%
, attributed to the W-O stretching modes. This band-like spectrum underlines
that all the as-deposited samples are amorphous \cite{Pezzoli_2015,
Baserga_2007, Bailini_2007, DiFonzo_2006}, although some differences
regarding band shapes and intensities can be seen in the spectra. Beside the
above bands, for deposition at pressure of 20 Pa or higher, the strong peak
of silicon substrate at 521 cm$^{-1}$ appears, revealing that such films are
transparent oxide. \newline
On the contrary, for O$_{2}$ pressure below 20 Pa,\ the Raman signal is
weak, and laser absorption is strong, such that the laser does not reach the
Si substrate. The Raman analysis thus confirms the semiquantitative results
of the electrical resistance measurements. This result is consistent with
the results of Yamamoto and coworkers \cite{yamamoto}, who see an abrupt
change of resistivity with varying oxygen content, consistent with the
transition from silvery to dark blue films. We therefore find that the broad
family of quasi amorphous W-O films, that have very similar structures,
witnessed by XRD spectra of the same type, can be subdivided in two groups.
Metal-like films are conductive and opaque, while oxide films with higher
oxygen content are non conductive and transparent. It has been
suggested \cite{bergreen} that the metal like behaviour can be due to a
semiconductive structure which is highly defective, such that many defect
induced electronic states lie in the gap; at higher oxygen content the
number of defects decreases and the band gap becomes observable. For the
sake of simplicity, in the rest of this paper the metal-like tungsten-oxygen
samples will be simply called metallic, or a-W(O), to underline their
amorphous nature, but with a significant oxygen content. On the other hand,
the transparent tungsten-oxide samples deposited at 20, 30 and 60 Pa are
called a-WO$_{3-x}$, where $3-x$ stands for possible stoichiometric defects.
All the spectra present an additional contribution, a band at about 960 cm$%
^{-1}$, attributed in literature to the stretching mode of the W = O bonds
at the surface of nanoclusters and void structures. It is thus related to
material nanocrystallinity and porosity \cite{Pezzoli_2015, DiFonzo_2006}.
This band is separated from the high frequency band only in the case of
optically transparent films, but unfortunately it superimposes to the second
order scattering of the Si substrate. Only at 60 Pa it evolves into a better
defined peak, confirming the high degree of nanostructuration and the high
surface-to-volume ratio of porous a-WO$_{3-x}$ coatings. 
\begin{figure}[h!]
\centering
\includegraphics[width = 0.5\columnwidth]{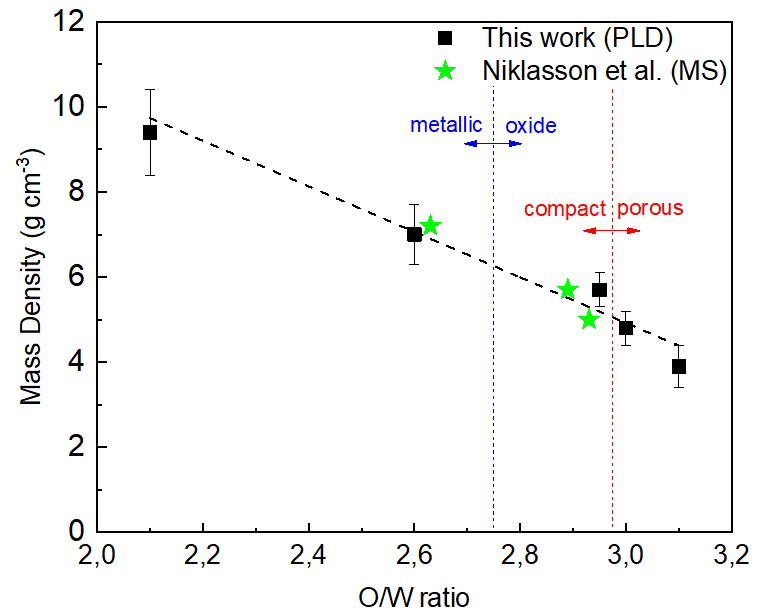}
\caption{Measured mass density as function of the O/W ratio. Measured data
are compared to literature values of WO$_{x}$ films deposited by
magnetro-sputtering \protect\cite{Niklasson}. The blue dotted-line separates
metallic a-W(O) samples from oxide a-WO$_{3-x}$ ones.}
\label{MassDensity}
\end{figure}
\newline
The mass density $\rho $ of the samples turns out to be strictly correlated
to the oxygen content. Figure \ref{MassDensity} shows the linear dependence
of $\rho $ on the O/W ratio. In the case of a-W(O), $\rho $ goes from 9.4 g
cm$^{-3}$ to 7 g cm$^{-3}$. For a-WO$_{3-x}$, instead, $\rho $ is $\approx $
5.6 g cm$^{-3}$ at O/W = 2.95, 4.8 g cm$^{-3}$ at O/W = 3 and 3.9 g cm$^{-3}$
at O/W = 3.1. In these cases, the obtained values are below the bulk value
of crystalline WO$_{3}$ (7.1 g cm$^{-3}$ \cite{Rottkay}), remarking the
higher porosity and amorphous structure that characterizes our samples. As a
comparison, we reported in figure \ref{MassDensity} the mass densities
related to sputtered WO$_{x}$ films that are commonly exploited for solar
cells research and electrochromic devices \cite{Niklasson}. As it can be
seen, sputtered and PLD films having similar O/W ratios also have very
similar mass densities. This is an important and somewhat surprising result
since the PLD and the sputtering processes are characterized by very
different energies of the ablated particles that, in turn, could deeply
affect the structure of the film and its mass density. Since it is well
known that $\rho $, disregarding the crystalline size, has a strong
influence on the thermomechanical properties of a material, one can expect
similar thermomechanical properties between PLD and sputtered films of
comparable O/W ratio. This would extend the results we obtain for PLD films
to a more general family of tungsten-oxygen coatings, the PLD process being
able to extend the range of accessible O/W ratios.

\subsection{Residual stresses and elastic moduli of as-deposited coatings}

\label{result2}

\begin{figure}[h!]
\centering \includegraphics[width = 0.5\columnwidth]{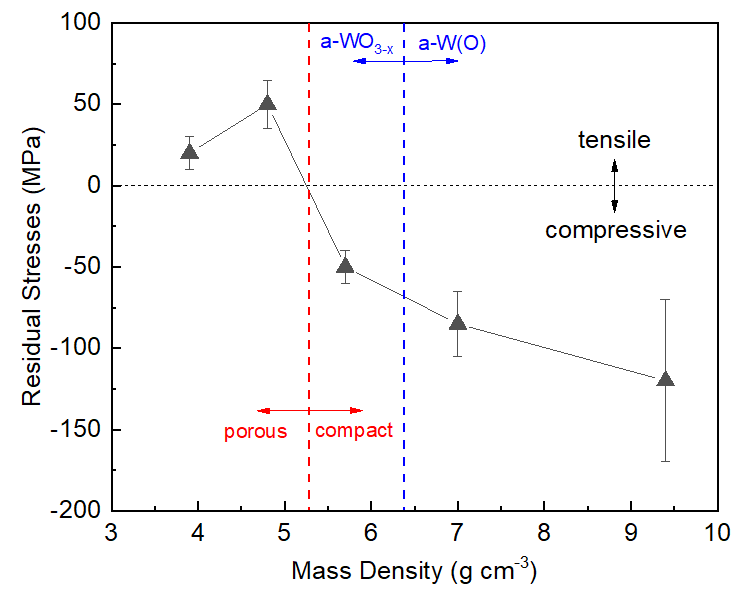}
\caption{Measured residual stresses as function of mass density. The red
line separates compact samples from porous ones, the blue one, instead,
a-W(O) from a-WO$_{3-x}$}
\label{sigma_res}
\end{figure}

The obtained results for the residual stresses $\sigma _{res}$ are shown in
figure \ref{sigma_res} as functions of the mass density $\rho $.\ Two main
regions can be observed, which correspond to the change of film morphology:
(i) compressive stresses in the case of compact samples, (ii) tensile
stresses in the case of porous films. Moreover,within the compressive region 
$\sigma _{res}$ depends on $\rho $, decreasing from -120 MPa in the case of
metallic a-W(O) to -50 MPa for a-WO$_{3-x}$, with $\rho $ ranging from 9.4
to 5.6 g cm$^{-3}$. On the other side, the tensile stress decreases from $%
\approx $ 50 MPa to 20 MPa as $\rho $ decreases from 4.8 to 3.9 g cm$^{-3}$.

The observed different nature of the residual stresses can be directly
associated to the PLD process, and more generally to the growth of $\sigma
_{res}$ in PVD coatings. In general, the magnitude and the nature of $\sigma
_{res}$ is associated to various contributions, namely intrinsic stresses ($%
\sigma _{i}$) and thermal stresses ($\sigma _{th}$) that arise during
deposition. In the particular case of PLD at room temperature intrinsic
stresses prevail over the thermal counterpart. A detailed description of the
nature of $\sigma _{i}$ can be found elsewhere \cite{Daniel_2010}. For the
purpose of the present work, it is important to underline the fact that
intrinsic stresses are determined by the sum of various contributions: (i)
tensile stresses originating from the grain growth process ($\sigma _{growth}
$), (ii) compressive stresses related to the adatoms diffusion to grain
boundaries ($\sigma _{diff}$) and (iii) compressive stresses related to ion
irradiation of the growing surface ($\sigma _{ion}$). Among them, $\sigma
_{ion}$ is commonly the dominant part when the energy of the ablated species
is sufficiently high to promote knock-on displacements of surface adatoms
(i.e. \textit{atomic peening effect}). This leads to a consequent formation
of defects and the growth of a compressive $\sigma _{i}$ that can reach up
to several GPa. In our case, the highest particles energy is found for
depositions at the lowest pressures, resulting in compact high density
coatings, consistently characterized by a compressive $\sigma _{res}$. When
the energy decreases, $\sigma _{ion}$ becomes less relevant. In the case of
porous films, the existing intercolumnar voids network limits atom mobility between columns. This inhibits
grain boundaries motion, such that the tensile $\sigma _{growth}$ prevails
on the compressive $\sigma _{diff}$. This is again in accordance with our
experimental evidence in the case of porous films, where $\sigma _{res}$ is
tensile. \newline
This well distinct $\sigma _{res}$ behavior can provide a guidance for the
selection of films for various applications. Indeed, a compressive residual
stress is beneficial for the mechanical behavior of the coatings, by
increasing the cracks, wear and corrosion resistance. Such a resistance is
particularly important for applications where high temperatures or external
loads are applied. Tensile stresses, on the contrary, tend to decrease
fatigue strength and life, to increase crack propagation, and to lower the
resistance to environmentally assisted cracking. The higher cracking
probability of tensile coatings is well highlighted by SEM images in figure %
\ref{SEM_asdep}, where a high density of through-thickness cracks is found
at 60 Pa. \newline
The elastic moduli of the coatings are then determined by Brillouin
spectroscopy. As discussed in section \ref{experimental2}, the moduli are
derived under the isotropic homogeneous condition. This condition is well
satisfied for compact amorphous coatings. However, tree-like nanostructure
in porous samples can induce a substantial anisotropy of the mechanical
properties of the films, such as the in-plane properties can differ from the
out-of-plane ones. Possible anisotropy effects on the elastic moduli of
nanostructured tungsten films measured by Brillouin spectroscopy have
already been considered \cite{Besozzi_2016}. The elastic moduli reported
here for porous samples, computed under the isotropic assumption, can be
seen as lower bounds for the real anisotropic moduli \newline
\begin{figure}[h!]
\centering \includegraphics[width = 0.5\columnwidth]{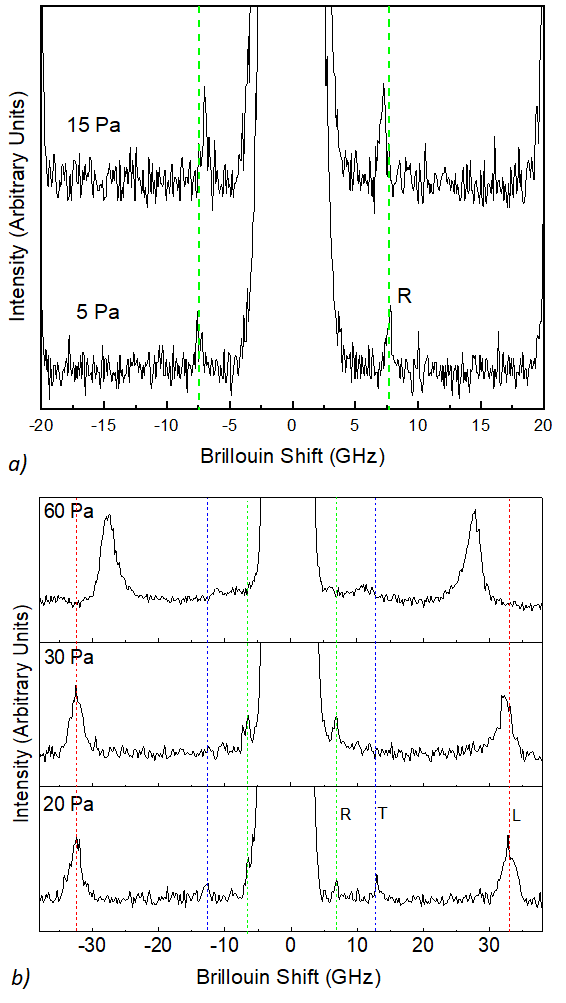}
\caption{Brillouin spectra recorded at an angle of incidence $\protect\theta $ = 60$^{\circ }$.
$R$ = Rayleigh mode, $T$ = transverse bulk mode, $L$ =
longitudinal bulk mode. a) a-W(O) coatings,  b)  a-WO$_{3-x}$ coatings}
\label{Brillo_asdep}
\end{figure}
Brillouin spectra recorded for a-W(O) and a-WO$_{3-x}$ samples are shown in
figure \ref{Brillo_asdep}a and \ref{Brillo_asdep}b respectively. In the case
of metallic a-W(O) coatings, only a low frequency mode can be detected. This
mode is associated to the surface Rayleigh wave ($R$) of the film, the
prototype of SAWs. In the case of optically transparent WO$_{3-x}$ two
additional peaks become visible: the mid frequency transverse bulk acoustic
wave ($T$) and the high frequency longitudinal bulk acoustic wave ($L$). The
spectra obtained at 20 and 30 Pa are quite similar, while at 60 Pa an
evident shift of the modes towards lower frequencies is detected. The
frequencies of these peaks depend on different factors, such as the elastic
properties, mass density and the refractive index, as mentioned in section %
\ref{experimental}. The observed peaks shifts can be thus attributed to a
variation of all these properties. In addition, at 60 Pa the $R$ mode
disappears. This is due to the open porous morphology, that does not support
surface waves propagation. 
\begin{table}[h]
\centering {\relsize{0.4} 
\begin{tabular}{lcccccc}
\toprule O$_2$ Pressure & O/W & $\rho$ & thickness & $n$ & $\sigma_{res}$ & 
Elastic modulus \\ 
(Pa) &  & (g cm$^{-3}$) & ($\mu$m) & (532 nm) & (MPa) & (GPa) \\ 
\toprule 5 & 2.1 & 9.4 & 3.1 $\pm$ 0.32 & - & - 120 $\pm$ 50 & 125 $\pm$ 20
\\ 
15 & 2.6 & 7 & 3.1 $\pm$ 0.35 & - & - 85 $\pm$ 25 & 74 $\pm$ 10 \\ 
20 & 2.95 & 5.7 & 3.3 $\pm$ 0.4 & 1.88 $\pm$ 0.1 & - 50 $\pm$ 5 & 72 $\pm$ 8
\\ 
30 & $\approx$ 3 & 4.8 & 3.3 $\pm$ 0.4 & 1.68 $\pm$ 0.08 & 50 $\pm$ 7 & 68 $%
\pm$ 5 \\ 
60 & 3.1 & 3.9 & 3.2 $\pm$ 0.35 & 1.49 $\pm$ 0.1 & 20 $\pm$ 5 & 43 $\pm$ 8
\\ 
\bottomrule &  &  &  &  &  & 
\end{tabular}
}
\caption{O/W stoichiometric ratio, mass density, thickness, refractive index
(at 532 nm), residual stress ($\protect\sigma _{res}$) and elastic modulus ($%
E$) of the coatings.}
\label{tab_n}
\end{table}

In the case of a-W(O) coatings, the thickness of the sample (i.e. $\approx $
3 $\mu $m) is such that the displacement field associated to SAWs is
essentially confined within the films. The coatings thus behave like 
semi-infinite media, with two main consequences. On one hand, the SAWs are
not dispersive (i.e. the frequency of the modes does not depend on the wvevector,
i.e. on $\theta $), such that, in principle, the information from only one
measurement at one incidence angle $\theta $ could be sufficient. On the
other hand, this limits the number of possible detectable SAWs. For a-W(O)
films, only the elastic information carried by the $R$ wave can be
exploited, following the procedure described in section \ref{experimental},
thus limiting the accuracy in the estimation of the elastic moduli. In the
case of a-WO$_{3-x}$ films, instead, the elastic moduli can be derived
directly from the bulk peaks frequencies through equations \ref{vL}, \ref{vT}
and \ref{bulkwaves}. In order to do that, the refractive index of the films
must be known. In this case, the simultaneous presence of the $R$ and $T$
waves, observed at 20 Pa and 30 Pa, is exploited to derive a consistent
estimation of $n$. Due to its predominantly shear nature, the Rayleigh
velocity $v_{R}$ can be approximated in terms of $v_{T}$ as $v_{R}\approx
f(\nu )v_{T}$ \cite{Freund_1998}, where $f(\nu )=\frac{0.862+1.14\nu }{1+\nu 
}$, $\nu $ being the Poisson's ratio of the film. In the case of W
based materials, the Poisson's ratio has been found in the range between
0.28 and 0.45 \cite{Besozzi_2016}. With $\nu $ in this range, the values of $%
f(\nu )$ remain between two close bounds: $0.933\pm 0.015$, such that
considering $v_{T}\approx 0.933v_{R}$ introduces at most a 1.6\% error in
the approximation of $n$. For this reason, we compute $v_{R}$ from the $R$
peak frequency (see fig. \ref{Brillo_asdep}b), and then we substitute it in
equation \ref{bulkwaves} to extract an estimation of $n$. This procedure is
done for the samples deposited at 20 and 30 Pa, where the $R$ and $T$ modes
are simultaneously present. At 60 Pa, the $R$ mode is not present. Since it
is reasonable to consider that the polarizability of a-WO$_{3-x}$ does not
change between 20 Pa and 60Pa (e.g. no effects related to crystallization), $%
n$ can be consistently estimated by means of the well known \textit{%
Lorenz-Lorentz} correlation \cite{LorentzLorentz}: 
\begin{equation}
\rho _{60}=\rho _{20}\frac{n_{20}^{2}-1}{n_{20}^{2}+2}\frac{n_{60}^{2}+1}{%
n_{60}^{2}-1}  \label{Lorentz-Lorentz}
\end{equation}%
where $\rho _{20}$ and $\rho _{60}$ are the mass densities of the amorphous
coatings at 20 and 60 Pa respectively, while $n_{20}$ and $n_{60}$ the
corresponding refractive indexes. The obtained values of $n$ at 532 nm are
summarized in table \ref{tab_n}. For example, in the case of compact a-WO$%
_{3-x}$ we obtain a refractive index of $\approx $ 1.88 which is in good
agreement with the ones reported in literature for compact amorphous WO$_{3}$
films (i.e. $\approx $ 1.9) \cite{Rottkay, Sun2010}. 
\begin{figure}[h!]
\centering \includegraphics[width = 0.5\columnwidth]{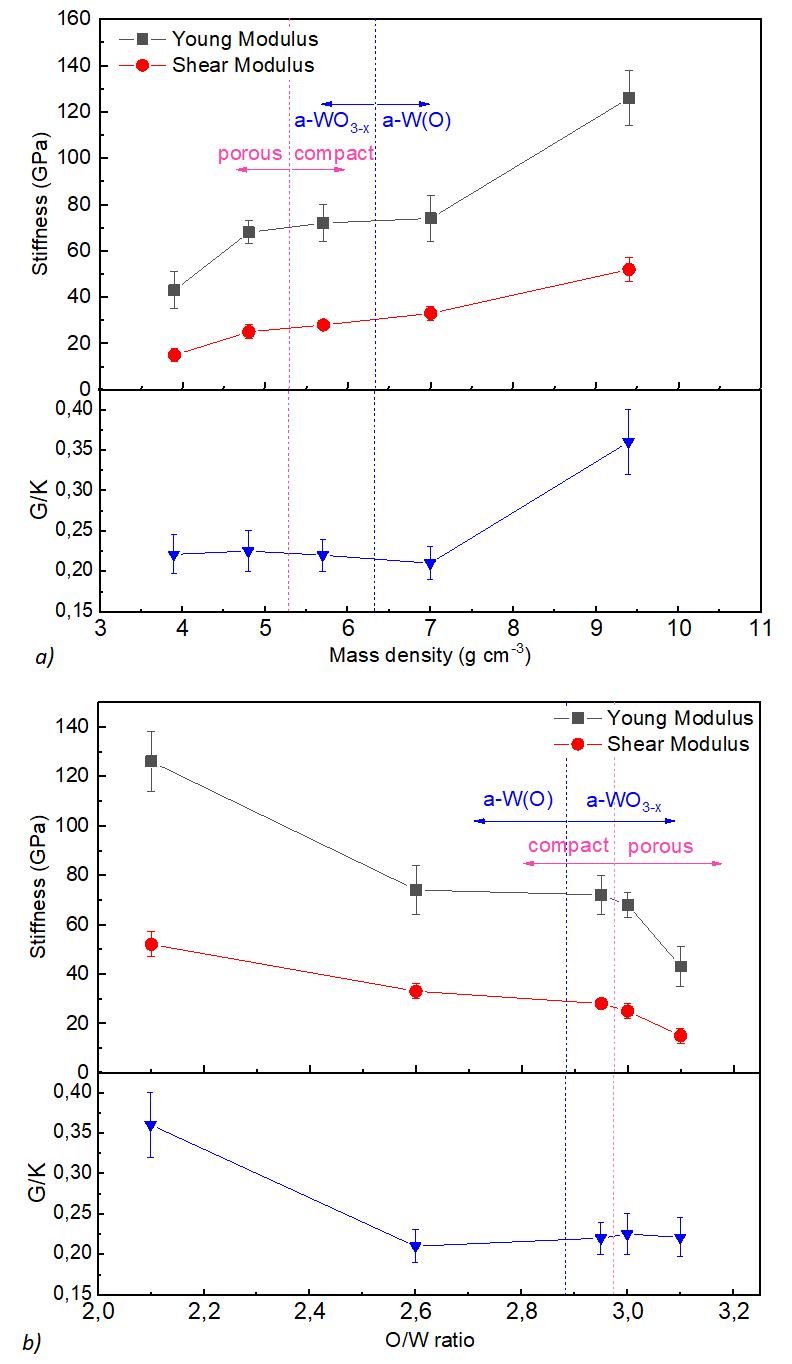}
\caption{Elastic moduli the coatings. a) Young Modulus $E$, shear modulus $G$
and $G/K$ ratio as function of film mass density. b) the same properties as
function of the stoichiometric ratio.}
\label{Mecc_asdep}
\end{figure}
\newline
From the estimation of $n$ at 532 nm, $C_{11}$, $C_{44}$ and all the elastic
moduli are finally obtained. Figure \ref{Mecc_asdep} summarizes the values
of the elastic Young modulus ($E$), the shear modulus ($G$) and the shear to
bulk modulus ratio ($G/K$), which, in particular, can be seen as an index of
local ductility of the material \cite{Besozzi_2016}. In figure \ref%
{Mecc_asdep}a the moduli are plotted as function of $\rho $, while in figure %
\ref{Mecc_asdep}b versus the O/W ratio. As it can be seen, a strong
dependence of the moduli on $\rho $ and the O/W ratio is found: as $\rho $
decreases and O/W increases the stiffness drastically decreases. Metallic
a-W(O) coatings are characterized by a higher stiffness with respect to the
oxide counterpart. In the metallic region $E$ goes from $\approx $ 125 GPa
to $\approx $ 74 GPa, coherently $G$ drops from 52 GPa to 33 GPa. $G/K$,
instead, goes from $\approx $ 0.36 to $\approx $ 0.21. This is a non-obvious
trend, since from SEM and Raman analysis there is not an appreciable
difference between these samples. At $\rho $ = 7 g cm$^{-3}$, which
corresponds to O/W = 2.6, a plateau of the moduli is reached: the material
changes its chemical configuration, by forming tungsten-oxide, but the
properties are not affected even if $\rho $ decreases between the samples.
In particular, $E$ remains between 68 and 72 GPa, $G$ between 25 and 28 GPa
and $G/K$ $\approx $ 0.21. This is consistent with the XRD analysis, which
indicates very similar structures of these films, while resistivity and
Raman measurements indicate instead a modification of the electronic
states. Finally, when evident intercolumnar pores appear, $E$ and $G$
further drop to $E$ = 43 GPa and $G$ = 16 GPa, at 60 Pa, which confirms that
mass density is the main parameter in determining the elastic moduli. 
\newline
It is interesting to compare the stiffness observed for metallic a-W(O)
films with the one proper of metallic amorphous-like W (i.e. $E$ = 150 GPa, $%
G$ = 50 GPa $\rho $ = 9 g/cm$^{3}$ \cite{Besozzi_2016}). For an O/W ratio of
2.1 the decrease is not so important (125 GPa) considering that the a-W(O)
system is determined by a different interplanar distance, compared with
amorphous-like W (respectively 3.54 $\mathring{A}$ and 2.31 \AA ), and an
amount of stored oxygen which is fivefold (2.1 compared with 0.4). Rising
the oxygen content O/W from 2.1 to 2.6, the material becomes much more soft
probably due to a decrease of the film density. Also the mechanical
properties of a-WO$_{3-x}$ are lower in comparison with cubic WO$_{3}$ (i.e. 
$E$ = 258 GPa, $G$ = 100 GPa, $G/K$ = 0.48 \cite{Liu2018}). In this case,
instead, the difference can be attributed to the specific amorphous
structure of the film. As a result of the loss of the long range order
proper of crystalline materials, the interatomic potential in the case of
amorphous materials can be lowered. This, in turn, can be associated to a
higher mean interatomic distance, which means a lower mean interatomic
binding energy, so lower elastic moduli \cite{Liang}. However, this can
confer the material some peculiar properties. For example, the drop observed
for $G/K$ is related to an increase of local ductility, so to a higher
ability of the material to locally allocate shear flow. In these terms, a-WO$%
_{3-x}$ films can be macroscopically brittle but microscopically capable of
sustaining shear flow \cite{Schuh}. \newline
These results can be compared to the few ones reported in literature for PVD
WO$_{3}$ coatings \cite{Carrejo_2016, Parreira_2006, Polcar_2006}. Parreira
et al. \cite{Parreira_2006} found a Young modulus of amorphous WO$_{3}$
films of $\approx $ 100 GPa, which is slightly higher than our values. For
lower oxygen contents (i.e. O/W below 3) the discrepancy is less
pronounced: they found $E$ varying between 170 and 100 GPa for O/W ratios
between 2 and 2.6, which corresponds to the $E$ values we measured for
a-W(O) coatings of 125 GPa and 74 GPa. Polcar et al. \cite{Polcar_2006} and
Carrejo et al. \cite{Carrejo_2016}, instead, report higher Young modulus
(i.e. between 110 and 164 GPa) for compact amorphous WO$_{3}$ coatings.
Nevertheless, the correlation between all these results is difficult since
there is no information about the material mass density, that, as already
mentioned, severely affects the elastic properties of the material. However,
the authors show an evident softening of the coatings with O$_{2}$
enrichment in the deposition atmosphere, which is in agreement with our
results. \newline
The previous mechanical characterization highlighted that compact a-WO$_{3-x}
$ films (deposited at 20 Pa of O) are characterized by interesting
mechanical properties, such as compressive residual stresses and high local
ductility, which can be fruitful for a wide range of applications. For this
reason, they are chosen as the reference samples for successive
characterizations. We thus characterize the $CTE$ of as-deposited a-WO$_{3-x}
$ films and the evolution of $\sigma _{f}$ during thermal treatments, as
well as we investigate the influence of different annealing temperatures on
the morphology, the structure and on the elastic moduli.

\subsection{Coefficient of thermal expansion and stress evolution of a-WO$%
_{3-x}$ coatings}

\begin{figure}[h!]
\centering \includegraphics[width = 0.5\columnwidth]{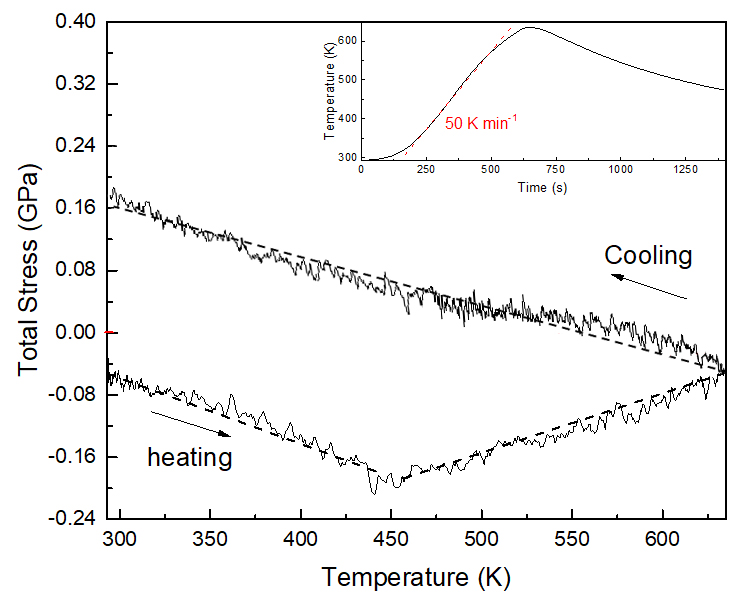}
\caption{Film stress evolution during annealing evaluated by the SC method.
The black dotted line represents a linear fit of the mean value of the
stress.}
\label{stresstemp}
\end{figure}

The $CTE$ of a-WO$_{3-x}$ films is determined by the procedure described in
section \ref{experimental2}. The standard thermal treatment adopted to this
purpose is shown in the inset of figure \ref{stresstemp}; the total film
stress $\sigma _{f}$ is monitored during heating and cooling. The linear fit
of the mean value of the stress (i.e. the dotted line) during the first
heating steps is then exploited to derive the $CTE$ of the film. From 
equation \ref{StoneyTh}, the slope of the dotted line is equal to $\frac{%
d\sigma _{f}}{dT}=\frac{E}{1-\nu }(\alpha _{f}-\alpha _{Si})$. In this way,
we obtain a mean $CTE$ of 8.6 $\cdot $ 10$^{-6}$ K$^{-1}$ between 273 - 450
K. Very little information about the CTE of tungsten-oxide is available in
literature, but it is known that tungsten-oxide materials can have very
different thermal expansion coefficients, that are strongly correlated with
the specific stoichiometry and can even be negative at high temperatures.
For example, the CTE at room temperature can vary from 1.3 $\cdot $ 10$^{-6}$
K$^{-1}$ in the case of W$_{18}$O$_{49}$ to 3.3 $\cdot $ 10$^{-6}$ K$^{-1}$
for WO$_{2}$ and between 8 and 15 $\cdot $ 10$^{-6}$ K$^{-1}$ for WO$_{3}$ 
\cite{West, Tokunaga, Takamori}. The value we find is higher than the one of
the Si substrate (i.e. 2.6 $\cdot $10$^{-6}$ K$^{-1}$), which explains the
negative slope of the linear fit, and is in agreement with the range of
values reported in literature. Differences between literature values and our
result can be attributed to a different structure of the material (e.g.
crystalline vs amorphous) and possible stoichiometric defects. \newline
As it can be seen in figure \ref{stresstemp}, a compressive residual stress
is initially developed during the heating cycle for temperatures up to 450
K. The stress magnitude increases, as expected, linearly with heating
temperature up to 450 K. However, after 450 K, which corresponds to a stress
of $\approx $ -180 MPa, the stress variation reverses its sign, the total
stress eventually becoming tensile. This is a strong evidence of the
beginning of some structural evolution. The tensile nature of the developed
stress can be associated to the observed volume shrinkage which, in turn,
relates to the beginning of diffusion and grain growth processes \cite%
{Daniel_2010}. These processes continue until the maximum temperature is
reached. Upon cooling, instead, the tensile stress associated to the
coefficient of thermal expansion mismatch between the film and the substrate
increases linearly with decreasing temperature. The linear trend upon
cooling shows almost the same slope of the heating cycle. This slope, as
already mentioned, is related to the $CTE$ and the elastic modulus of the
coating. The slope upon cooling almost equal to the one upon heating
is an indication that no irreversible changes of
these macroscopic properties have been induced by the fast thermal
treatment. The kinetics of structural evolution is probably slower than
the total annealing time. What changes is the value of $\sigma _{f}$. After
the thermal cycle, the film is found in a tensile state of stress. This
underlines that $\sigma _{res}$ relaxation has occurred, due to some local
structure reorganization associated to defects diffusion at relatively low
temperatures.

\subsection{Thermal annealing of a-WO$_{3-x}$ coatings}

\begin{figure}[h!]
\centering \includegraphics[width = 0.5\columnwidth]{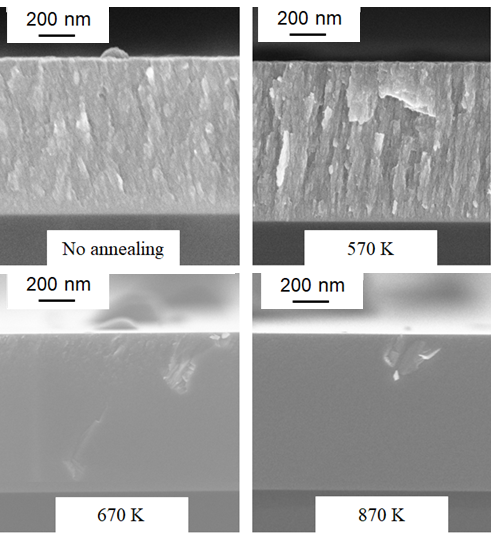}
\caption{SEM images of a-WO$_{3-x}$ samples annealed  for 1 hour in vacuum at
different temperatures.}
\label{SEM_annealed}
\end{figure}
Temperature induced effects on a-WO$_{3-x}$ films is investigated by vacuum
annealing treatments performed between 570 K and 870 K. SEM cross section
images of annealed a-WO$_{3-x}$ coatings are shown in figure \ref%
{SEM_annealed}. Up to 570 K no substantial morphological changes are
visible. Starting from 670 K, instead, SEM analysis clearly shows a
morphology modification. The compact nanostructured morphology of
as-deposited a-WO$_{3-x}$ evolves into a featureless, more compact one. This
can be related to the annealing driven structural reorganization process,
which, in turn, suggests that around 670 K crystallization occurs. SEM cross
section images also show a decrease by about 20\% of film thickness starting
from 670 K. The thickness, indeed, goes from 3.3 $\mu $m to $\approx $ 2.7 $%
\mu $m at 670 K. This, in turn, is associated with an increase of film mass
density, that goes from the as-deposited value of 5.7 g cm$^{-3}$ to $%
\approx $ 7 g cm$^{-3}$ above 670 K. 
\begin{figure}[h!]
\centering \includegraphics[width = 0.5\columnwidth]{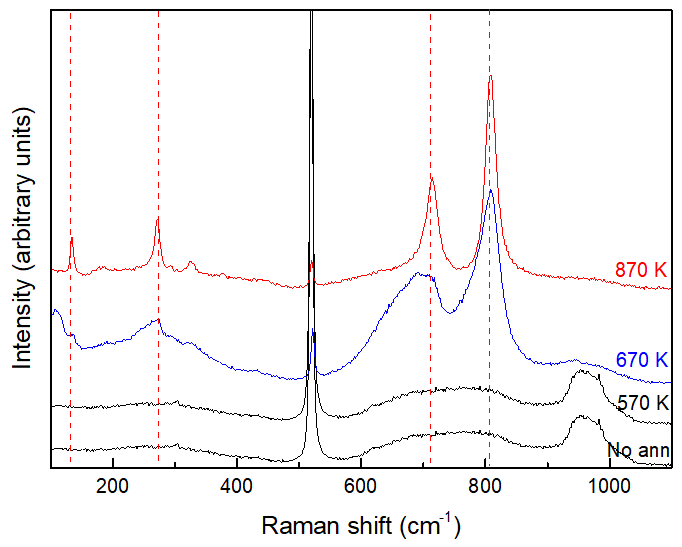}
\caption{Raman spectra of annealed a-WO$_{3-x}$ samples. The red dotted
lines correspond to the principal peaks proper of $\protect\gamma $
monoclinic WO$_{3}$. Si substrate peaks are present at 521 cm$^{-1}$ and at
960 cm$^{-1}$.}
\label{Raman_annealed}
\end{figure}
\newline
Raman spectroscopy is exploited to better highlight the observed
crystallization process. The obtained spectra are shown in figure \ref%
{Raman_annealed}. Up to 570 K, no remarkable differences of the spectrum
with respect to the as-deposited case can be detected. This is in agreement
with SEM analysis, which does not report a strong variation of the film
structure and morphology. At 670 K, instead, several new peaks become
visible. These peaks grow in correspondence of 133 cm$^{-1}$, 273 cm$^{-1}$,
715 cm$^{-1}$ and 805 cm$^{-1}$, which correspond to the principal peaks of
the crystalline $\gamma $ monoclinic phase of WO$_{3}$ \cite{Hayashi}. The
observed peaks width, in particular the width of the peak at 715 cm$^{-1}$,
suggests that, at this temperature a consistent amount of amorphous
structure is still present: the crystallization process is not completed. At
870 K, instead, definitely sharper peaks are found. Moreover, the positions
of these peaks are slightly shifted with respect to the ones of $\gamma $-WO$%
_{3}$. This can be attributed to different factors, such as internal
stresses developed during annealing or stoichiometry defects. For the former
case, as highlighted by stress evolution measurements of figure \ref%
{stresstemp}, the growth of a new crystalline phase in an amorphous matrix
can lead to the development of a high internal state of stress associated to
atoms diffusion, grains coalescence and growth. For the latter, EDXS
analysis performed on annealed samples confirm a slight reduction of the O/W
ratio for a-WO$_{3-x}$ from $\approx $ 2.95 down to 2.88. This could
suggest that a small part of the total amount of O$_{2}$ is only trapped in
the as deposited film and can desorb at relevant temperatures. 
\begin{figure}[h!]
\centering \includegraphics[width = 0.5\columnwidth]{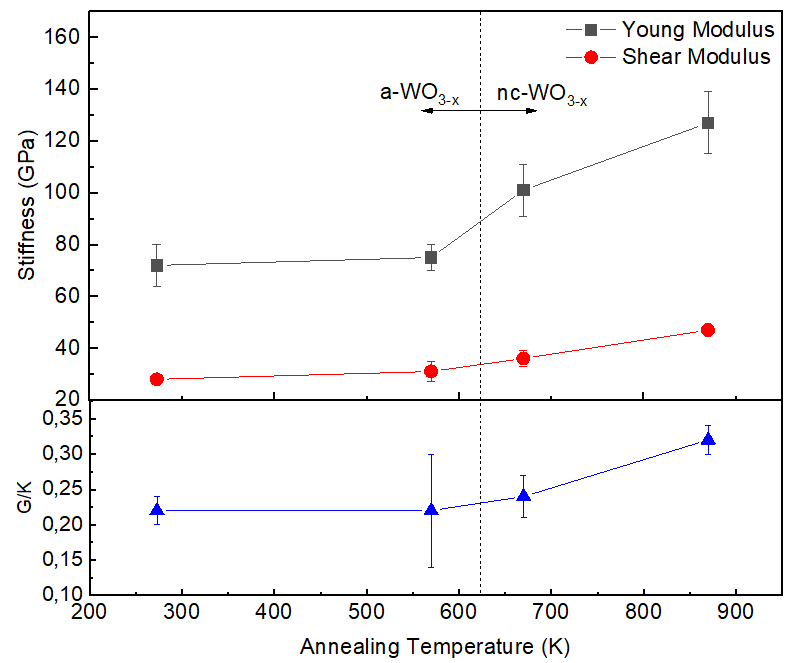}
\caption{Elastic moduli of annealed a-WO$_{3-x}$ films.}
\label{Mecc_annealed}
\end{figure}
\newline
The observed crystallization affects the mechanical properties of the films,
which are assessed by Brillouin analysis following the same procedure
adopted for a-WO$_{3-x}$ samples. The
obtained elastic moduli are summarized in figure \ref{Mecc_annealed}. As
expected, if annealing is performed at 570 K only a slight increase of
material stiffness is observed. At 670 K, instead, when crystallization
begins, $E$ = 101 GPa, $G$ = 36 GPa and $G/K$ = 0.24, with a consequent
increase by about 36\% of $E$ with respect to the as-deposited condition. In
agreement with what we found in \cite{Besozzi_2016},these properties can be
due to an amorphous matrix in which nanocrystals are embedded. Finally, when
the $\gamma $ monoclinic WO$_{3}$ phase is better defined at 870 K, a
further increase of the moduli is detected (i.e. $E$ = 127 GPa, $G$ = 50 GPa
and $G/K$ = 0.38). These results are in accordance with what we observed in
figure \ref{stresstemp} concerning the total stress evolution. We determined
the relaxation temperature $T_{R}$ to be at 450 K, and we found that even if
we rapidly heat the sample at temperatures $T>T_{R}$ no changes of the
macroscopic properties are observed. This agrees with the slight stiffening
observed at 570 K. Consistently, only 1 h annealing treatments at 670 K
can alter significantly the nanostructure and the properties of the
material. The corresponding stiffening process is, in turn, attributed to
different competitive processes, such as structural relaxation, free volume
annihilation and short-range ordering (i.e. nano-crystals formation), which
can increase the interatomic potential and consequently the elastic moduli
of the material \cite{Ouyang}.

\section{Conclusions}

\label{Conclusions}

In this work we investigate the thermomechanical properties of different
systems of amorphous tungsten-oxygen and tungsten-oxide coatings. Such
properties are important for the design and construction of devices, in
which the tungsten based layer is generally in contact with other layers,
and in which the mechanical integrity is a crucial requirement. This is
particularly critical when the films are required to operate at high
temperatures or in the presence of external loads (e.g. thermophotovoltaic,
solar cells, electronics) The samples were deposited by PLD, which allowed
us to access a wide range of morphologies, structures and compositions of
the films. We thus found the relationship between these properties and the
thermomechanical ones. We explored a whole interval of properties, 
offering useful information to identify the type of coating which best fits 
each single application. \newline
The mechanical properties of the as-deposited films, namely the elastic
moduli and the residual stress, resulted to be simultaneously influenced by
the morphology and the O/W ratio: as the films become less dense and rich of
oxygen, the stiffness and residual stresses linearly decrease. However, for $%
\rho $ between 4.8 - 7 g cm$^{-3}$ and O/W ratio between 2.6 and 3 the
stiffness undergoes only small variations, consistently with a structure,
revealed by XRD, which remains almost the same, while the electronic
properties, revealed by resistivity and Raman measurements, have more
significant variations. Among tungsten-oxide coatings, compact a-WO$_{3-x}$
films showed the most promising mechanical properties in the as deposited state, with a
moderate stiffness, compressive residual stresses and a relatively high
ability to locally allocate shear flow. These properties can be desired for 
applications under external thermal or mechanical loads. The quite high CTE
and a quite low relaxation temperature must be considered for their use in
high temperature applications. Crystallization into the monoclinic $\gamma $%
-WO$_{3}$ phase starts at 670 K, with a consequent strong crystallization
induced stiffening and a decrease of local ductility. The lower observed
relaxation temperature (i.e. 450 K) suggests that possible structural
relaxation and diffusion processes already begin at temperatures well below
the determined annealing temperature, without affecting the overall
properties of the material. Since the knowledge of annealing kinetics can be
crucial to determine the evolution of films properties under high
temperatures, this type of analysis will require further investigation.
Overall, the ensemble of our results can provide a guidance in the design of
various devices which exploit tungsten based layers.

\section*{Acknowledgments}

This work has been carried out within the framework of the MISE-ENEA `%
\textit{Accordo di Programma}' (AdP), PAR2015 and PAR2016. The research
leading to these results has also received funding from the European
Research Council Consolidator Grant ENSURE (ERC-2014-CoG No. 647554). The
views and opinions expressed herein do not necessarily reflect those of the
European Commission.

\section*{Data availability}
The raw data required to reproduce these findings cannot be shared at this time due to technical or time limitations.
The processed data required to reproduce these findings cannot be shared at this time due to technical or time limitations.

\bibliographystyle{model1a-num-names}
\bibliography{<your-bib-database>}

\end{document}